\documentclass{article}

\usepackage[accepted]{icml2023}
\usepackage{microtype}
\usepackage{graphicx}
\usepackage{subfigure}
\usepackage{booktabs} 

\usepackage{hyperref}

\usepackage{icml2023}

\usepackage{amsmath}
\usepackage{amssymb}
\usepackage{mathtools}
\usepackage{amsthm}

\usepackage[capitalize,noabbrev]{cleveref}

\theoremstyle{plain}

\theoremstyle{definition}

\theoremstyle{remark}

\usepackage[textsize=tiny]{todonotes}

\icmltitlerunning{Submission and Formatting Instructions for ICML 2023}

\begin{document}

\twocolumn[
\icmltitle{DMOps: Data Management Operations and Recipes}



\icmlsetsymbol{equal}{*}

\begin{icmlauthorlist}
\icmlauthor{Eujeong Choi}{yyy,equal}
\icmlauthor{Chanjun Park}{yyy,equal}
\end{icmlauthorlist}

\icmlaffiliation{yyy}{Upstage, Gyeonggi-do, Republic of Korea}

\icmlcorrespondingauthor{Chanjun Park}{chanjun.park@upstage.ai}

\icmlkeywords{Machine Learning, ICML}

\vskip 0.3in
]



\printAffiliationsAndNotice{\icmlEqualContribution} 

\begin{abstract}
Data-centric AI has shed light on the significance of data within the machine learning (ML) pipeline. Recognizing its significance, academia, industry, and government departments have suggested various NLP data research initiatives. While the ability to utilize existing data is essential, the ability to build a dataset has become more critical than ever, especially in the industry. In consideration of this trend, we propose a "Data Management Operations and Recipes" to guide the industry in optimizing the building of datasets for NLP products. This paper presents the concept of DMOps which is derived from real-world experiences with NLP data management and aims to streamline data operations by offering a baseline.
\end{abstract}

\section{Introduction}
With the emergence of Data-centric AI~\cite{polyzotis2021can,mazumder2022dataperf}, various in-depth natural language processing (NLP) data research has been introduced in academia alongside the wide range of policies from industry and government departments~\cite{pencheva2020big}.

In the case of academia, there are studies boosting model performance through large-scale datasets~\cite{liu2021large,costa2022no} along with the production of benchmark datasets for objective performance comparison between models~\cite{wang2018glue,ruder2021challenges}. Furthermore, there are also benchmark datasets that specialize in specific tasks~\cite{rajpurkar2016squad,alt2020tacred}. The government contributes to the field by implementing public data open policies and releasing datasets from the National Statistics department~\cite{panagos2012european}.

However, the industry frequently dives into an untapped and specialized domain, where a ready-to-go dataset is rarely there. Especially for B2B companies, there is usually an urgent demand for data that fits their customers or business items~\cite{pustejovsky2012natural}. Since the open source and benchmark datasets are normally insufficient to meet these specific demands, additional data production is always a necessary step to initiate a particular task. As a result, the majority of AI businesses started to build their own task-specific datasets, alongside the emergence of companies that specialize in operating crowd workers to meet these demands. Additionally, research on efficient data production on human-in-the-loop started to make appearance~\cite{doan2018human,wu2022survey}.

Despite its necessity, there has been a paucity of studies in the field of NLP data production from an industry perspective. To the best of our knowledge, there has not yet been research that proposes the entire process starting from analyzing the business standpoint to data annotation and evaluation. Therefore, we propose a "Data Management Operations and Recipes (DMOps)" that will assist in building an NLP dataset efficiently and economically. Specially, we propose a DMOps that can produce high-quality NLP data needed in manufacturing deep learning models.

\section{DMOps}
Data management operations involve the integration of human input and decision-making into a data management pipeline or system. This involves a series of tasks such as data annotation, data quality assurance, and other activities that require a human touch~\cite{doan2018human, goosen2019system, solis2019human, eo2021comparative}. One way to implement data management operations is through the use of recipes. Recipes are step-by-step instructions for performing a specific task or set of tasks and can be used to guide human workers through the NLP data management process.

Our Data Recipes consists of 12 steps. Through these steps, we go over the entire process of data operations : from establishing the goal of the project to delivering the final data to the modeling team. The name and explanation of each step is as follows.

\paragraph{1. Establish the Project Goal}: Analyzing the purpose and requirements of data production is the first step of the recipe. This step requires collaboration with a team of NLP engineers and business operators. Through communication, we can decide the input and output format of data that is suitable to the model of choice, and also set data milestones that fit the timeline of the business operation team. 

Unlike academia where research starts from related works or enhancing existing benchmark datasets, the industry starts with its users and customer needs~\cite{tarafdar2019using, kerzel2021enterprise, alenezi2022can}. This gap between the two areas must be considered when setting the goal in the first place. To build a good dataset in the industry, we need to start from the end-user's needs and requirements~\cite{laato2022explain}.
    
\paragraph{2. Secure Raw Data}: Researching and collecting raw data is the second step of the recipes. Five possible cases of collecting raw data are 1) the client providing the raw data~\cite{ahmed2004applications}, 2) using open-sourced public data~\cite{lhoest2021datasets}, and 3) purchasing the raw data from its source platform, 4) crawl from website~\cite{pant2004crawling} 5) crowdsourcing~\cite{estelles2012towards, hossain2015crowdsourcing}. The key issue here is the copyright of each data source. License information must be checked thoroughly, and getting a legal review is recommended before its usage~\cite{khayyat2015open}.

Furthermore, providing access and editing rights to the raw data must be dealt with caution. To prohibit unauthorized modifications from multiple parties, a sturdy data storage structure with limited access rights are necessary. 

\paragraph{3. Data Pre-processing}: The third step is improving the quality of the raw data through pre-processing. Basically the pre-processing consists of two main tasks: first, adjusting the format of data regarding its requirements, second, filtering non-ethical, privacy invading, and noisy data~\cite{wiegand2018overview,park2020quality}. This step is all about practicing quality over quantity.

These preprocessing steps can be broadly divided into two tasks. The first task is to improve the quality of the data based on the inherent characteristics of the data~\cite{rahm2000data, ridzuan2019review}. A representative example of this is parallel corpus filtering in machine translation~\cite{koehn2020findings}, which is a data-centric method that improves the performance of the model without changing the structure of the model by only controlling the quality of the data~\cite{park2022asr}.

The second task is to address ethical issues with the data~\cite{van2004ethical}. This includes attaching data license information in advance or masking personal information if it exists. If these tasks are not clearly carried out in advance, there may be a risk of not being able to use the data even after data annotation and validation~\cite{martin2020ethical}. This is why it is considered a highly important task.
    
\paragraph{4. Design a Data Schema}: The fourth step involves designing an efficient annotation schema that captures all required information. This step is crucial, as it requires capturing the desired information fully while also ensuring efficiency to prevent cognitive overload for annotators. Also, figuring out parts that can be somewhat automated (pseudo-labeling) and parts that need human intervention (annotating) is essential in making the process efficient and moreover, accurate. With few pilot annotation iterations, the data scheme is expected to reach its optimal design~\cite{gregor2020research, he2022generate}.

This step is one of the most important steps in designing data, such as "What kind of meaningful information to extract in Information extraction (IE)?~\cite{hobbs2010information}", "What entity to tag in name entity recognition (NER)~\cite{mansouri2008named}", "In document summarization How much information should be compressed?~\cite{yao2017recent}", "When constructing data for machine translation (MT), should paraphrase, literal translation, or transcendental translation be used?~\cite{stahlberg2020neural}".

In academia, these elements are already pre-determined and research is conducted in that state, but in industry, this information must also be re-designed according to customer needs~\cite{tseng2021sustainable}.

\paragraph{5. Prepare a Guideline}: The fifth step is the documentation of the data scheme. Its purpose is to deliver the designed annotation system to the expected annotators. The difficulty of the guideline should be monitored since the clarity and detailed explanation may be in a trade-off relationship.
    
\paragraph{6. Recruit Annotators}: The sixth step is recruiting the annotators. The key is to select workers that are fit for the task for an efficient and accurate outcome. The best case would be selecting those who scored high on a test similar to the actual annotation task. Additionally, ethical considerations are also necessary. Good data is one that is created with fair compensation for the workers and without any unnecessary costs. It is important to take these factors into account as well~\cite{foley2014contingent}.
    
\paragraph{7. Instruct Annotators}: The seventh step is instructing the annotators with the guideline made above. In this stage, two-way communication that draws out questions and debates is the key whereas one-sided communication is discouraged.
    
\paragraph{8. Data Annotation}: This step involves annotating the actual data, where annotators transfer their linguistic, cognitive, and visual intuition into the data. To ensure consistency among annotators, it is important to establish a unified approach and provide a safe environment for questions. Keeping a question log is also suggested to prevent inconsistencies within the corresponding answers.

In addition, in the field of NLP, the development of annotation tools is of paramount importance. This is due to several key factors, including the need for quality control, efficiency, and scalability~\cite{grosman2020eras}. In terms of quality control, an annotation tool allows for data annotation to be performed in a consistent and accurate manner. This is crucial for ensuring the quality of the data used for training NLP models. Additionally, an annotation tool can make the annotation process more efficient, which is especially important when building large datasets or for data that needs to be annotated quickly~\cite{pei2022potato}.

Scalability is also a crucial factor to consider when developing an annotation tool. As the amount of NLP data continues to grow, a tool that can handle the volume of data is essential. Furthermore, an annotation tool can be designed to assist human annotators, providing suggestions and feedback, thus increasing annotation speed and reducing human error~\cite{perry2021lighttag}.

To sum up, the development of a well-designed annotation tool is essential for NLP data management. It is a crucial step in training and evaluating NLP models and can greatly aid in preserving and analyzing linguistic data.
    
\paragraph{9. Data Internal Factor Verification}: This ninth step is inspecting the annotated data. During this step, inspectors, who are usually selected from the annotator pool, must identify commonly occurring human errors and sort out the edge cases through discussions. Considering the nature of the Human-in-the-loop process, this step is essential to ensure the fidelity of the dataset.

In this stage, we recommend consensus labeling~\cite{tang2011semi}. Inter-annotator agreement (IAA) should be used to check the consistency of data labels~\cite{ragheb2013inter, bohavc2017text} due to the potential for human errors and misunderstandings of annotation guidelines. As human annotators are prone to inconsistencies and errors, it is necessary to verify the consistency of labels through IAA.

Certainly, data annotation can also be carried out in two steps. In the first round, rough annotation is performed on about 10 data, and IAA is checked in advance to correct the workers' misunderstandings.  In the second round, using the corrected knowledge of the workers, the task of annotating all data is undertaken in a more thorough manner. In other words, by effectively combining the data annotation stage and the data inspection stage, the efficiency of the workers can be improved.

To sum up, data internal factor verification step is a process of validation for the inherent elements of the data. However, it does not verify external factors or the relationship between the data and the model (i.e. if the data truly helps improve the performance of the model). Therefore, additional steps such as "data external factor verification" and "data evaluation through model verification" should be considered to fully validate the data.

\paragraph{10. Data External Factor Verification}: The tenth step is verifying the data external Factor. When inspecting data, it is necessary to first determine whether the work has been completed by observing the given guideline. Also, 1) data sufficiency, 2) data diversity, 3) data trustworthiness, 4) data privacy and security 5) data ethics suitability should be reviewed~\cite{roh2019survey,koo2022k}. 

In other words, going beyond the internal information of the data, it is a step to thoroughly examine the sufficiency, diversity, reliability, security, and ethics of the data from various perspectives. The best way to conduct this verification is through Institutional Review Boards or external advisory committees~\cite{blee2011ethics}.
    
\paragraph{11. Data Evaluation via Model Verification}: The eleventh step is verifying the quality of data through actual modeling. In order to quantitatively verify whether the data is made as planned, various experiments are conducted such as checking data efficiency by increasing the amount of data or sectioning the data to check the consistency of its quality~\cite{moon2021filter,park2021should}. It is natural to find artifacts within one's data; after identifying the repeated errors, revisiting the recipes from step 5 is frequently required to enhance the quality of data. If there are parts that do not match our purpose while proceeding the steps, we should return to stage 5 and revise the guideline for another iteration.

To complement the 'human-in-the-loop' cycle, it's essential to detect errors through the model and clean them through human intervention. This cycle ensures error-free data and alignment with the model's outcomes, resulting in cohesive results. The goal is to create data that is both error-free and coherent with the model.

\paragraph{12. Data Deliverables}: Final step of the recipes is delivering the final data outcome. In other words, it is the process of delivering annotated data to engineers or customers. When delivering, the versioning must be adapted to the protocol, and it is important to reveal the features of the data with its sample. Furthermore, after going through the exploratory data analysis (EDA) process, it is recommended to deliver the data analysis and the quality evaluation document together.

The quality of data in industry can be assessed based on several factors, including the informativeness of metadata, the legitimacy of data sources (e.g., compensated workers without unnecessary cost), well-established versioning systems, and intuitive and organized data storage structures. While these factors may seem obvious, they are crucial to creating high-quality data. In academia, these factors may not be given as much weight, but in industry, they are critical and together can elevate good data to the level of great data.

\section{Discussion of DMOps}
\begin{figure}[htp]
    \vspace{0.05cm}
    \centering
    \includegraphics[width=\columnwidth]{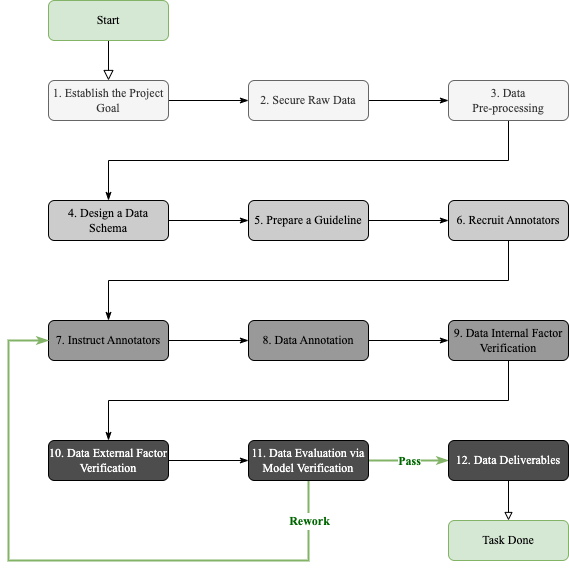}
    \caption{Image of the Data Management Operation Recipes}
    \label{fig:Data Pipeline}
\end{figure}

\paragraph{Why DMOps?} Our proposed "DMOps" provides a universal and fixed process for data construction that can be applied to any NLP task or domain. This approach can serve as a baseline for data production, as it ensures a consistent and reliable process for generating high-quality data. The visualization of DMOps is shown in Figure~\ref{fig:Data Pipeline}.

\paragraph{Future of DMOps}
As DMOps evolve, automation will play a larger role in data production. This requires improving the efficiency and automation of current human-performed tasks. Additionally, synthetic data will become increasingly important, and methods for efficient inspection of this data through self-annotation by humans need to be developed. Managing data generated by large-scale language models like ChatGPT~\footnote{\url{https://chat.openai.com/}} and GPT-3~\cite{brown2020language} efficiently is crucial in maintaining high data quality.

\section{Conclusion and Future Works}
In this paper, we presented DMOps, a task-agnostic methodology for efficiently producing high-quality NLP data with human annotation, which can serve as a baseline for any NLP data production. To increase the reliability of the proposed process, we plan to conduct quantitative verification at each stage of the process in the future. Additionally, we aim to conduct a study to investigate the difference in data quality when using the proposed data recipes compared to not using them.

\section*{Acknowledgements}

\nocite{langley00}

\bibliography{example_paper}
\bibliographystyle{icml2023}

\end{document}